\documentstyle[12pt]{article}
\newcommand{\version}{\mbox{1.2}}
\newcommand{\fortran}{\mbox{\sc Fortran}}
\newcommand{\lambdams}{\mbox{$\Lambda_{\overline{\mbox{\tiny MS}}}$}}
\newcommand{\dach}{\mbox{$\hat{\hphantom{x}}$}}
\newcommand{\docuname}{\mbox{\bf  DISJET}}

\newcommand{\oasz}{\mbox{$\mbox{O}(\alpha_{s}^{2})$}}

\def\Lms#1{\Lambda_{\overline{MS}}^{(#1)}}

\def\ams{\alpha_{\overline {MS}} }
\def\thebibliography#1{{\bf{References}}\list
 {[\arabic{enumi}]}{\settowidth\labelwidth{[#1]}\leftmargin\labelwidth
   \advance\leftmargin\labelsep
   \usecounter{enumi}}
   \def\newblock{\hskip .11em plus .33em minus -.07em}
   \sloppy
   \sfcode`\.=1000\relax}
  
\begin{document}
\thispagestyle{empty}
\noindent
\hbox to \hsize{
\hskip.5in \raise.1in\hbox{\bf University of Wisconsin - Madison}
\hfill$\vcenter{\hbox{\bf MAD/PH/821}
            \hbox{April 1994}}$ }

\vspace{.5in}
\begin{center}
  \begin{Large}
  \begin{bf}
 \docuname \hspace{2mm}\version\hspace{2mm}: \\
A Monte Carlo Program for Jet Cross Section\\ Calculations in
 Deep Inelastic Scattering\\
  \end{bf}
  \end{Large}
  \vspace{0.8cm}
  \begin{large}
{T. Brodkorb$^a$ and E. Mirkes$^b$}\\[1cm]
  \end{large}
{\it
$^a$
Laboratoire Physique Theorique et Hautes
Energies, Universit{\'e} Paris-Sud, 91405 Orsay,
 France.
Address after June 1993, SAP Walldorf, Germany\\
\vspace*{3mm}
$^b$Physics Department, University of Wisconsin, Madison WI 53706, USA\\[2cm]
}
{\bf Abstract}
\end{center}
\begin{quotation}
\noindent
We present a new parton level Monte Carlo  program for the
calculation of jet cross sections in Deep Inelastic Scattering based on
Born and next-to-leading order matrix elements.
Using a class of  invariant jet definition schemes,  the program allows
 for the calculation of differential distributions of jet cross sections
in the basic kinematical
 variables (like $s, x, y, W^2, Q^2\ldots$) as well as for  total jet
cross sections.
Various kinematical cuts can be chosen from an input file.
\end{quotation}
\newpage
\section{Introduction}
The start-up of the HERA electron-positron collider
 marked  the beginning of a new aera of experiments exploring
 Deep Inelastic Scattering (DIS) of electrons and protons.
One of the topics to be studied at HERA will be the Deep
 Inelastic ($\equiv$ high $Q^2$) production of multi jet
events, where good event statistics are expected
allowing for precision tests of QCD.

In this paper, we present a Monte Carlo program  \docuname\
that allows for the calculation of next-to-leading order (NLO)
 $(1+1)$,$(2+1)$ and leading order (LO)
 $(3+1)$
jet cross sections\footnote{$''+1''$ denotes the  remnant jet} in DIS.
Using a class of invariant jet definition scheme  \cite{herai}
 the program allows
 for the calculation of differential distributions of jet cross sections
in the basic kinematical
 variables as well as for  the calculation of total jet cross sections.

The source code is written in \fortran\ and running ability is
tested on a DEC-Station
 and an ALPHA-Station  running  the Operating Systems VMS
and  on a Sun-Sparc station running under UNIX.
The Monte Carlo routines are using the  VEGAS-package \cite{vegas}
for
numerical integration of functions depending on up to ten variables.
Parton distributions are incorporated from the
PDFLIB-package \cite{pdflib}.
Each ($n+1$) jet cross section in \docuname\ is calculated in
fixed order QCD perturbation theory. Therefore a comparison of
jet-measurements with our predicted QCD-results provides a direct tool
of determining $\alpha_s$ or \lambdams. \\[3mm]
{\bf Changes to \docuname\ 1.0}\\[2mm]
An error in the finite part of the
\oasz\ virtual corrections to the quark initiated subprocess
is corrected.
An error in the labeling of the \oasz\
(3+1)-jet subprocesses [{\it iproc} = 1001,1002,1003]
is corrected.
An additional cut in the variable
$z= \frac{pp_1}{pq}$, where $p_1$ is the four momentum
of one of the final partons in (2+1) jet production
is implemented.

\section{Matrix Elements}
In \docuname\ complete fixed order QCD matrix elements
for the process
\begin{equation}
e^-(l) + \mbox{proton}(P) \rightarrow \mbox{proton remnant}(p_r) +
\mbox{parton} \,\,1 (p_1)
\ldots
+\mbox{parton}\,\, n (p_n)
\label{form1}
\end{equation}
$(n=1,2,3)$ are used.  Reaction (\ref{form1}) proceeds via the exchange of an
intermediate vector boson $V=\gamma^\ast, Z,W$. At present only the
exchange by a virtual photon is incorporated in
\docuname.
We denote the
$\gamma^{\ast}$-momentum by $q$, the absolute square by $Q^2$, the
center of mass energy by s, the square of the final hadronic mass by
$W^2$ and introduce the scaling variables $x$ and $y$:
\begin{eqnarray} \label{defkin} q & = & l-l' \nonumber \\
Q^2 & \equiv & -q^2=xys>0 \nonumber \\
s & = & (P+l)^2 \nonumber \\ W^2 & \equiv & P_f^2=(P+q)^2 \\
x & = & \frac{Q^2}{2Pq} \hspace{1cm} (0<x \le 1) \nonumber \\
y & = & \frac{Pq}{Pl} \hspace{1cm} (0<y \le 1) \nonumber \end{eqnarray}
At fixed s, only two variables in (\ref{defkin}) are independent, since e.~g.
\begin{displaymath} xW^2=(1-x)Q^2,\hspace{1cm}  Q^2=xys. \end{displaymath}
Averaging over the azimuthal angle between the parton plane and the
lepton plane $(\vec{l},\vec{l'})$ (in the
($\gamma^\ast$-initial parton)-cms)
 the jet cross sections for the
exchanged virtual photon factorize the following $y$ dependence \cite{herai}:
\begin{equation}
d\sigma[n-\mbox{jet}] =
\frac{2\pi\alpha^2}{xyQ^2}
\left\{ (1+(1-y)^2)\,d\sigma_{U+L}[n-\mbox{jet}]
-y^2\,d\sigma_{L}[n-\mbox{jet}]\right\}
\label{form2}
\end{equation}
In eq. (\ref{form2}) $\sigma_{U+L}[n-\mbox{jet}]$ and $\sigma_L[n-\mbox{jet}]$
 denotes the cross section
contributions from an unpolarized and longitudinal polarized gauge boson.
These helicity cross sections
 are linearly related to polarization density matrix elements
of the virtual boson. One has:
\begin{equation}
\sigma_U=\frac{1}{2}(h_{++}+h_{--})
\hspace{1cm}
\sigma_L=h_{00}
\hspace{1cm}
\sigma_{U+L}\equiv\sigma_U+\sigma_L
\end{equation}
where
$
h_{mm^{\prime}}= \epsilon_{\mu}^{\ast}(m)H^{\mu\nu}\epsilon_{\nu}(m^{\prime})
$, ($m,m^{\prime}=+,0,-$) and $\epsilon_{\mu}(\pm) (\epsilon_{\mu}(0))$ are the
transverse (longitudinal) polarization vectors of the $\gamma^\ast$ in the
($\gamma^{\ast}-$initial parton)-cms.
The helicity cross sections
can technically
 be obtained by the following projections on the (partonic) hadrontensor
$H^{\mu\nu}$, which is calculated in fixed order perturbation theory,
\begin{eqnarray}
\hat{\sigma}_{U+L}&=&\left(-\frac{1}{2}g_{\mu\nu}+\frac{3x_p}{pq}p_{\mu}p_{\nu}\right)
\,\,H^{\mu\nu}[n-\mbox{jet}]
\label{ul}\\
\hat{\sigma}_{L}&=&\frac{2x_p}{pq}p_{\mu}p_{\nu}\,\,H^{\mu\nu}[n-\mbox{jet}]
\end{eqnarray}
where $p$ denotes the momentum of the initial state parton and $x_p=Q^2/(2pq)$.

The available jet multiplicities
in \docuname\ are listed below:
\begin{itemize}
\item[1)]
(1+1) jet \hspace{1cm} $O(\alpha_s^0)$ and $O(\alpha_s^1)$
\item[2)]
(2+1) jet \hspace{1cm} $O(\alpha_s^1)$
\item[3)]
(2+1) jet \hspace{1cm} $O(\alpha_s^2)$
\item[4)]
(3+1) jet \hspace{1cm} $O(\alpha_s^2)$
\end{itemize}
The physics of 1) and 2) is extensively discussed in \cite{herai}.
For numerical results see also \cite{heraii,heraiii}.  Matrix elements of the
{\it complete} contributions to (2+1) jet  $O(\alpha_s^2)$
corrections are first discussed in \cite{tomzpc}.
In \docuname, a modified version of these matrix elements is implemented
including the full NLO scale dependence.
NLO numerical results for (2+1) jet cross sections and rates
(based on \docuname)
including the contributions from all helicity cross sections
are presented in \cite{heraiii}.
The  (2+1) jet NLO
contributions originating
from the  projection with the metrical tensor $-g_{\mu\nu}$
(see eq. (\ref{ul}))  on the
hadron tensor and a discussion
 of the results are given in \cite{dirk}.
  Results for the
$O(\alpha_s^2)$ (3+1) jet cross sections are discussed in detail in
\cite{heraii}.

The general structure of the hadronic jet cross sections within the
framework of perturbative QCD is given by
\begin{equation}
d\sigma^{had}[(n+1)-\mbox{jet}] =
\int d\eta \,\,f_a(\eta,M_f^2)\,\,\, d\hat{\sigma}^a(p=\eta P,
\alpha_s(M_r^2))
\label{had}
\end{equation}
where one sums over $a=q,\bar{q},g$. $f_a(\eta,M_f^2)$
is the probability density to find a parton $a$ with fraction $\eta$
in the proton if the proton is probed at a scale $M_f^2$.
$\hat{\sigma}^a$ denotes the partonic cross section
from which collinear initial state
singularities have been factorized out  (in the NLO calculation)
at a scale $M_f^2$.
Note that  $\hat{\sigma}^a$ depends in NLO
also explicitly on the renormalization scale $M_r$ and factorization scale
$M_f$.
\docuname\ allows the user
to choose the  mass scales $M_r$ and $M_f$ in eq. (\ref{had})
 according to the following
input variables $a_x, b_x$ and $c_x$:
\begin{equation}
M_x^2 = a_x \,Q^2 \,+\,b_x\,W^2\,+\,c_x\,p_T^2  \label{mxdef}
\end{equation}
where $x=f,r$ (see also next section).
In eq. (\ref{mxdef}) $p_T$ denotes the transverse momentum of the two partonic
jets in $(2+1)$ jet production.
The transverse momentum is defined with respect to
the $\gamma^\ast$ direction.
$c_x \neq 0$ can only be chosen in the case of
 (2+1) jet production, otherwise $c_x$ is set to zero.
Furthermore, to  avoid to small scales for perturbation theory, the
renormalization and factorization scales are cut at 2 GeV$^2$, i.e.
\begin{equation}
M_x^2 = \mbox{max}\{ 2\, \mbox{GeV}^2, M_x^2\}
\end{equation}

In order to calculate the $(n+1)$ jet cross sections we have to {\it
define} what we call $(n+1)$ jets by introducing a {\it resolution
criterion}.
As has been elaborated in detail in \cite{herai,heraii} energy angle
cuts are not suitable for an asymmetric machine with its strong boosts
from the hadronic cms to the laboratory frame.
As a jet resolution criterion we use the invariant mass cut criterion
introduced  in
\cite{herai,heraii} such that
\begin{equation}
s_{ij}=(p_i+p_j)^2 \geq M^2=\max\left\{ y_{cut} M_c^2,M_0^2\right\}
\hspace{1cm}
(i,j=1,\ldots ,n,r;\hspace{5mm} i\neq j)
\label{jetdef}
\end{equation}
where $y_{cut}$ is the resolution parameter which should be chosen in
[0.01 \ldots 0.08] (see fig. 5 in \cite{heraii}).  $M_c$ is a
typical mass scale of the process which defines the jet definition scheme.
  This mass scale can be chosen by
the user according to  the values of $a_c,b_c$ and a constant $C$ [GeV$^2$]:
(see also  section 3)
\begin{equation}
M_c^2= a_c Q^2+ b_c W^2 + C
\end{equation}
The extreme choices $a_c=0 (b_c=0)$ and $C=0$ correspond to the
$''W-\mbox{scheme}''
\,\,\, (''Q-\mbox{scheme}'')$
as introduced in \cite{herai}.
$M_0$ in eq. (\ref{jetdef})
is an additional fixed mass cut which we have introduced in
order to clearly separate the perturbative and non-perturbative regime
in the case, where $M_c^2$ is small (i.e. $Q^2$ and/or $W^2$ and $C$ small).
A reasonable choice for $M_0$ is $M_0=2$ GeV \cite{heraii}.


\section{Input parameters}
The input parameters for \docuname\ are written to a file
referred to as unit 4 by the local \fortran\ compiler, i.e.\
for004.dat on VMS and fort.4 on UNIX.

Parameters to choose are: (an example for the input file
 is given in the appendix)
\begin{itemize}
\item {\it srs}: real, f6.2, [GeV]\\  center of mass energy $\sqrt{(P+l)^2}$
\item {\it ias}: 
integer, i1\\
choose one-loop ore two-loop formula for $\ams$.
The value of $\alpha_s$ is matched at the thresholds $q=m_q$
and the number of flavours $n_f$ in $\alpha_s$ is
 fixed by the number of flavours
for which the
 masses are less than $M_r$.
 Furthermore the numbers of quark flavours that can be
pair-produced are set equal to $n_f$ chosen in $\alpha_s$.
\begin{itemize} \item[1:] 1-loop-formula \item[2:] 2-loop-formula
\end{itemize}
\item {\it ilambda}:  $\{0,1\}$ integer, i1\,\,\,
\begin{itemize}
 \item[0:] $\Lms{4}$ in $\alpha_s$ is
           chosen consistent to the value in the chosen parton distribution
           functions (see below).
 \item[1:] $\Lms{4}$ in $\alpha_s$  can be chosen by hand (=$dlam$),
            see next item.
\end{itemize}
\begin{itemize}
\item {\it dlam}: real, f5.4, [GeV]\\
               only relevant in the case $ilambda=1$, otherwise this
                  number is changed in the main program.\\
                  If $ilambda=1$ ,  $\Lms{4}$=$dlam$, real, f5.4.,
                  [GeV].
\end{itemize}
\item {\it iaem}: $\{0,1\}$ integer, i1\\
\,\,\,fixed or $Q^2$ dependent
 $\alpha_{electromagnetic}$.
\begin{itemize}
 \item[0:] $\alpha=0.00729735$ \hspace{2mm} fixed
 \item[1:] $Q^2$ dependent $\alpha$
\end{itemize}
\item {\it icross}: $\{0,1\}$ integer, i1
 \begin{itemize} \item[0:] calculate the  jet cross section
as defined in (\ref{form2}) in [pb]
 \item[1:] only effective for iproc $<$ 1000 (see below)\\
calculates the helicity cross sections $\sigma_{U+L}[(n+1)-\mbox{jet}]$ or
$\sigma_{L}[(n+1)-\mbox{jet}]$ for fixed $x$ and $Q^2$ $(n=1,2)$
\item {\it x}:
         $[0.0001,1]$: real, f5.4\\
 If {\it icross}=0, $x$ is  changed by the integration routine.
\item {\it Q\dach 2}: $[4GeV^2,s]$: real, f10.2\\ same  as {\it x}
 \end{itemize}
\item {\it ipola}: $\{10,11,12\}$: integer, i2\\
 choose helicity cross sections in the
case of (n+1) jet production ($n=1,2$). Depending on $icross$, the full
cross sections (including all coefficients)
or the helicity cross sections $\sigma_{U+L, L}$
(for fixed $x,Q^2$)
 are calculated.
 \begin{itemize}
\item[10:]
take sum of all helicity cross sections, i.e.  calculates
$d\sigma[n-\mbox{jet}]$ as
given in (\ref{form2}) (only possible for $icross=0$).
\item[11:] calculates only the term $\sim
d\sigma_{U+L}[n-\mbox{jet}]$ in (\ref{form2})
\item[12:] calculates only the
term $\sim  d\sigma_{L}[n-\mbox{jet}]$ in (\ref{form2})
 \end{itemize}
\item {\it iproc}: $\{10,100,101,102,190,191,192,1001,1002,1003\}$:
integer, i4\\
      characterization of the process
\begin{itemize}
\item[10:] (1+1) jet $\equiv \sigma_{tot}$ in $O(\alpha_s^0)$
 \item[100:] (2+1) jet: complete, i.e. sum of quark and gluon initiated
processes
 \item[101:] (2+1) jet: quark initiated
\item[102:] (2+1) jet: gluon initiated
\item[190:]  $O(\alpha_s^1)$-contribution to $\sigma_{tot}$ in the DIS-scheme:
complete
\item[191:]  $O(\alpha_s^1)$-contribution to $\sigma_{tot}$ in the DIS-scheme:
quark initiated
\item[192:]  $O(\alpha_s^1)$-contribution to $\sigma_{tot}$ in the DIS-scheme:
gluon initiated
\item[1001:] (3+1) jet: $q\rightarrow q G G$ (class D \cite{heraii})
\item[1002:] (3+1) jet: $G\rightarrow G q \bar{q}$ (class C \cite{heraii})
\item[1003:] (3+1) jet $q\rightarrow q q^{\prime} \bar{q^{\prime}}$ (class E
\cite{heraii})
 \end{itemize}
\item {\it istruc}: $\{0,\dots,7$\}: integer, i1\\
       only relevant if {\it iproc=} $100,101,102$ \begin{itemize}
\item[0:] complete NLO corrections, (note: {\it ias} should be set equal to  2)
\item[1:] Leading order ($O(\alpha_s)$)
\item[2:] NLO virtual corrections  (finite part)
 \item[3:] NLO final state real
corrections
\item[4-7:] NLO initial state real corrections:
\\ 4: delta-part 5: hard-part 6: plus-part
7: substracted-part \\
(note: {\it ias} should be set = 2 for {\it istruc}=2-7)
 \end{itemize}
\item {\it y\_c}: $[0.01,0.1]$: real, f5.4\\ jet resolution parameter
$y_{cut}$ as defined in eq. (\ref{jetdef})
\item {\it M\_0\dach 2}: real, f3.1, [GeV$^2$]\\
 additional fixed jet cut mass as defined
in eq. (\ref{jetdef})
\item {\it M\_f\dach 2 = $a_f$ Q\dach 2 + $b_f$ W\dach 2} + $c_f\,\, p_T^2$:
  the  factorization mass in eq. (\ref{had}) can be chosen by the three
values $a_f,b_f$  and $c_f$ in a format f4.2 ($c_f\neq 0$ only effective  for
$iproc=100,101,102$)
\item {\it M\_r\dach 2 = $a_r$ Q\dach 2 + $b_r$ W\dach 2} + $c_r\,\, p_T^2$:
the  renormalization mass in eq. (\ref{had}) can be chosen by  the three
values $a_r, b_r$ and $c_r$ in a format f4.2
($c_r\neq 0$ only effective  for
$iproc=100,101,102$)
\item {\it M\_c\dach 2 = $a_c$ Q\dach 2 + $b_c$ W\dach 2} + $C$ [GeV]$^2$:
  the jet cut mass as defined in eq (\ref{jetdef})
can be chosen by the three values $a_c,b_c$
 and $C$ in a format f4.2
\item {$p^2_{T\,min}$}: real, f6.2\\
minimal required transverse momentum  for  the jets
(transverse with respect to the
 $\gamma^\ast$ direction). This entry is  effective only
 for $iproc=100,101,102$.

\item {$z_{min},z_{max}$} [0.,1.]: real, f8.2\\
 Cut on the invariant $z= \frac{pp_1}{pq}$, where $p_1$ is the four momentum
 of one of the final partons in (2+1) jet production.
 This entry is  effective only
 for $iproc=100,101,102$.

\item {\it Nptype, Ngroup, Nset}: integer, i3,i3,i3\\
      choose parton distribution functions according to \cite{pdflib}.
\item {\it iacc}: $\{1,2,3\}$ integer, i1\\
some predefined settings for the accuracy
       the VEGAS-integration: \\
       1: acc = 0.05,\hspace{5mm} ncall = 20000\\
       2: acc = 0.005,\hspace{5mm}ncall =  200000\\
       3: acc = 0.0005,\hspace{5mm}ncall = 1000000\\
\item {$ x_{min}, x_{max}$}: $[0.0001,1]$ real, f8.6, Integration limits
\item {$ y_{min}, y_{max}$}: $[0,1]$ real, f8.6, Integration limits
\item {$ Q^2_{min}, Q^2_{max}$}: $[4,s]$ real, f8.2, [GeV$^2$],
      Integration limits
\item {$ W^2_{min}, W^2_{max}$}: $[5,s]$ real, f8.2, [GeV$^2$],
            Integration limits
\item The next integer (format i2) determines the number of single
differential cross sections to be calculated.
\item the following input values are data-cards for the calculation of single
 differential cross sections (see the last column).
The user may change the entries in the first two columns. The first (second)
column defines the lower (upper) kinematical limit of the
variable specified in the last column.
The user may also change the single differential cross sections
to be calculated by suitable  changings in the subroutine
dsigma.

\end{itemize}



\section{Output}
The output of \docuname\ is written to the standard output as
well as to a file unit 98. The output file contains all input
values  specified in the unit 4 input file and all results
calculated by VEGAS, i.e. the integrated cross section and the
differential cross sections.

The differential cross sections will be written to file unit 98 in the
following format.
\begin{itemize}
\item the (up to) 10
      differential cross sections are written to the
      file one after another
\item every package of data belonging to one differential cross section
 is preceded by  the string given in the data
cards of the input file
\item each differential cross section is presented line by line for
      each requested bin in the form:\\ mean value of the variable in  the bin,
differential cross section in this bin, number of points in this bin.

\end{itemize}
\section{Summary and conclusion}
We present the first version of the Monte Carlo program \docuname.
 The program allows to calculate NLO (1+1) and (2+1) jet
cross sections and LO (3+1)
 jet cross sections in DIS. The contributions from all helicity
cross sections are included.
Note that
\docuname\ is not an event generator like  LEPTO  or HERMES \cite{lepto}.
Furthermore, the freedom to choose a jet definition scheme is restricted to the
schemes
discussed in the text and does not include the jet definition proposed in
\cite{catani}.

Some improvements and extensions are planed for further versions of the
program.
We would appreciate  comments and suggestions on the program to be send
to  mirkes@phenom.physics.wisc.edu.\\[5mm]
{\bf Acknowledgements:}\\
It is a pleasure to thank I.H. Park
and J. Rathsman
for their efforts to test the MC program \docuname\
 and helpful suggestions concerning
some extensions of the program.
This work is supported in part by the U.S. Department of Energy under
contract No. DE-AC02-76ER00881, and in part by the University of Wisconsin
Research Committee with funds granted by the Wisconsin Alumni Research
Foundation.\\[5mm]
\noindent
\def\npb#1#2#3{{\it Nucl. Phys. }{\bf B #1} (#2) #3}
\def\plb#1#2#3{{\it Phys. Lett. }{\bf B #1} (#2) #3}
\def\prd#1#2#3{{\it Phys. Rev. }{\bf D #1} (#2) #3}
\def\prl#1#2#3{{\it Phys. Rev. Lett. }{\bf #1} (#2) #3}
\def\prc#1#2#3{{\it Phys. Reports }{\bf C #1} (#2) #3}
\def\pr#1#2#3{{\it Phys. Reports }{\bf #1} (#2) #3}
\def\zpc#1#2#3{{\it Z. Phys. }{\bf C #1} (#2) #3}
\def\ptp#1#2#3{{\it Prog.~Theor.~Phys.~}{\bf #1} (#2) #3}
\def\nca#1#2#3{{\it Nouvo~Cim.~}{\bf #1A} (#2) #3}

\newpage
\noindent
{\bf Appendix}
\begin{verbatim}
======================================================
======= Example for an input-file ====================
======    for004.dat under VMS    ====================
======      fort.4   under UNIX   ====================
======================================================
SRS=     295.
IAS=     1
ILAMBDA= 0, DLAM= .1970
IAEM=    1
ICROSS=  0, X= .010, Q^2=  10000.00
IPOLA=   10
IPROC=   100
ISTRUC=  1
Y_C=     .0200, M_0^2= 4.0 GEV^2
M_F^2= 1.00 Q^2 + 0.00 W^2 + 0.00 PT^2
M_R^2= 1.00 Q^2 + 0.00 W^2 + 0.00 PT^2
M_C^2= 0.00 Q^2 + 1.00 W^2 + 0.00
PT2MIN=  0.
NPTYPE =  1, NGROUP =  3, NSET = 31
IACC=    1
XMIN= 0.001000 XMAX= 0.100000
YMIN= 0.040000 YMAX= 0.950000
ZP_MIN =   0.00   ZP_MAX =   1.00
Q^2_MIN=  10.00   Q^2_MAX= 87025.00
W^2_MIN= 600.00   W^2_MAX= 87025.00
10
-3.,    -0.1,  32,0,0,'LG(X)'
3.D-4, 1.D-1,  32,0,0,'X'
2.,    10.D0,  32,0,0,'SQRT(Q^2)'
10.,    1.D5,  32,0,0,'Q^2'
3.,   200.D0,  32,0,0,'SQRT(W^2)'
10.,    1.D5,  32,0,0,'W^2'
-2.,    -0.1,  32,0,0,'LG(Y)'
0.05,   0.95,  32,0,0,'Y'
0.67,   0.99,  32,0,0,'THRUST'
0.  ,    50.,  32,0,0,'PT'
0
0


 \end{verbatim}
\end{document}